\documentclass[a4paper]{article}
\title{DOUBLY SPECIAL RELATIVITY: A NEW RELATIVITY OR NOT?}
\author{Nosratollah Jafari$^1$
  ~~and~~  Ahmad Shariati$^2$
\\[5pt] $^1$ \textit{Institute for Advanced Studies in Basic Sciences,}
\\[0pt] \textit{P.O. Box 159, Zanjan 45195, Iran}\\
\\[5pt] $^2$ \textit{Department of Physics, Alzahra University,}
\\[0pt] \textit{Tehran 19938-91167, Iran.}\\
} 
\date{February 19, 2006}

\begin{document}
\maketitle
\begin{abstract}
\vspace{0.5cm} Double Special Relativity theories are the
relativistic theories in which the transformations between inertial
observers are characterized by two observer-independent scales of
the light speed and the Planck length. We study two main examples of
these theories and want to show that these theories are not the new
theories of relativity, but only are re-descriptions of Einstein's
special relativity in the non-conventional coordinates.
\end{abstract}
\vspace{0.5cm}

\section{Introduction}
\vspace{0.5cm}It seems that the Planck length $l_{p}$ has a crucial
role in the Quantum Gravity. In some scenarios of quantum gravity
like loop quantum gravity the Planck length or Planck scales act as
a threshold for quantum effects in the spacetime, beyond which the
usual description of spacetime breaks down. Thus, it seems that the
value of $l_{p}$ must have the same value in all inertial frames
and this is in conflict with Einstein's Special
Relativity\cite{KG,Am2}. Doubly Special Relativity( DSR) has
proposed for solving this puzzle \cite{MS1,Am1,Am2}.

The Magueijo-Smolin (Ms) \cite{MS1,MS2} and Amelino-Camelia
\cite{Am1,Am2} DSRs are two main examples of these theories that
take much attractions recently. Here, we want to investigate these
theories further.
\newpage

\section{Magueijo-Smolin DSR}
Let's now explain briefly the Magueijo - Smolin (Ms)
transformations: Magueijo and Smolin have looked for a non-linear
representation of the Lorentz group that remains the Planck length
as an invariant . If we denote the ordinary Lorentz boosts by
\begin{eqnarray}
L_{ab}=p_{a}\frac{\partial}{\partial p^{b}}-
p_{b}\frac{\partial}{\partial p^{a}},
\end{eqnarray}
then this representation can be obtained by using the modified
generators of boosts as
\begin{eqnarray}
K^{i}\equiv L_{0}^{i}+l_{p}p^{i}D.
\end{eqnarray}

Here, $D$ is the dilatation generator
              $$ D=p_{a}\frac{\partial}{\partial p_{a}}.$$

But, the rotation generators will be the unmodified $J^{i}\equiv
\epsilon^{ijk}L_{jk}.$ Also, the Lorentz Algebra remains intact:
$$ [J^{i},K^{j}]=\epsilon^{ijk}K_{k},~~~
[K^{i},K^{j}]=\epsilon^{ijk}J_{k},~~~~~~
[J^{i},J^{j}]=\epsilon^{ijk}J_{k}.$$

By acting the modified generators of boots on the momentum space, we
obtain the Magueijo - Smolin (MS) transformations between two
inertial systems which are in relative motion with constant speed
along the common $x$-axis as:
\begin{eqnarray}
\label{eq2a}
p_{0}'=\frac{\gamma(p_{0}-\frac{v}{c}p_{x})}{1+l_{p}(\gamma-1)p_{0}-l_{p}\gamma
\frac{v}{c}p_{x}},
\\ \label{eq2b}
p_{x}'=\frac{\gamma(p_{x}-\frac{v}{c}p_{0})}{1+l_{p}(\gamma-1)p_{0}-l_{p}\gamma
\frac{v}{c}p_{x}},
\\ \label{eq2c}
p_{y}'=\frac{p_{y}}{1+l_{p}(\gamma-1)p_{0}-l_{p}\gamma
\frac{v}{c}p_{x}},
\\ \label{eq2d}
p_{z}'=\frac{p_{z}}{1+l_{p}(\gamma-1)p_{0}-l_{p}\gamma
\frac{v}{c}p_{x}} ,
\end{eqnarray}

These transformations have many new features \cite {MS1,MS2}. For
example, they do not preserve the usual quadratic invariant on
momentum space. But, there is a modified invariant:
\begin{eqnarray}
\|p\,\|^{2}=\frac{\eta_{ab}p_{a}p_{b}}{(1-l_{p} p_{0})^{2}}
\end{eqnarray}
~~~~~~~~~~~~~~~~~~~~~~~~~~~~~~~~~~~~~~~~~~~~~~~~~~~~~~~~~~~~~~~~~~~~~~~~~~~~~~~~~~~~

Also, these transformations remain invariant the light speed "$c$"
and the Planck length "$l_{p}$" as desired. This property can be
seen from MS transformations and equation (7).

But, looking closer at these transformations, we can see that
changing 4-momentum $p_{\mu}$ to
                              $$\pi_{\mu}=\frac{p_{\mu}}{1-l_{p}p_{0}}$$
then the MS transformations become
\begin{eqnarray}
\left\{%
\begin{array}{ll}
    \pi'_{0}=\gamma (\pi_{0}-\frac{v}{c}\pi_{x})\\
    \\
    \pi'_{x}=\gamma (\pi_{x}-\frac{v}{c}\pi_{0}) \\
    \\
    \pi'_{y}=\pi_{y} \\
    \\
    \pi'_{z}=\pi_{z} \\
\end{array}%
\right.
\end{eqnarray}
These are the same ordinary Lorentz transformations for momentum
space. Therefore, \emph{the MS transformations are probably only
re-description of the usual Lorentz transformations in the
non-conventional coordinates} \cite{Jaf,AHL1}. This fact can also be
seen from the MS momentum space diagram. In this figure the MS
momentum $p_{0}$ and $p_{1}$ are drawn as horizontal and radial
lines.

\begin{figure}
\begin{center}
\begin{picture}(100,100)(0,0)
\multiput(0,0)(100,0){2}{\line(0,1){100}}
\multiput(0,0)(0,100){2}{\line(1,0){100}} \includegraphics{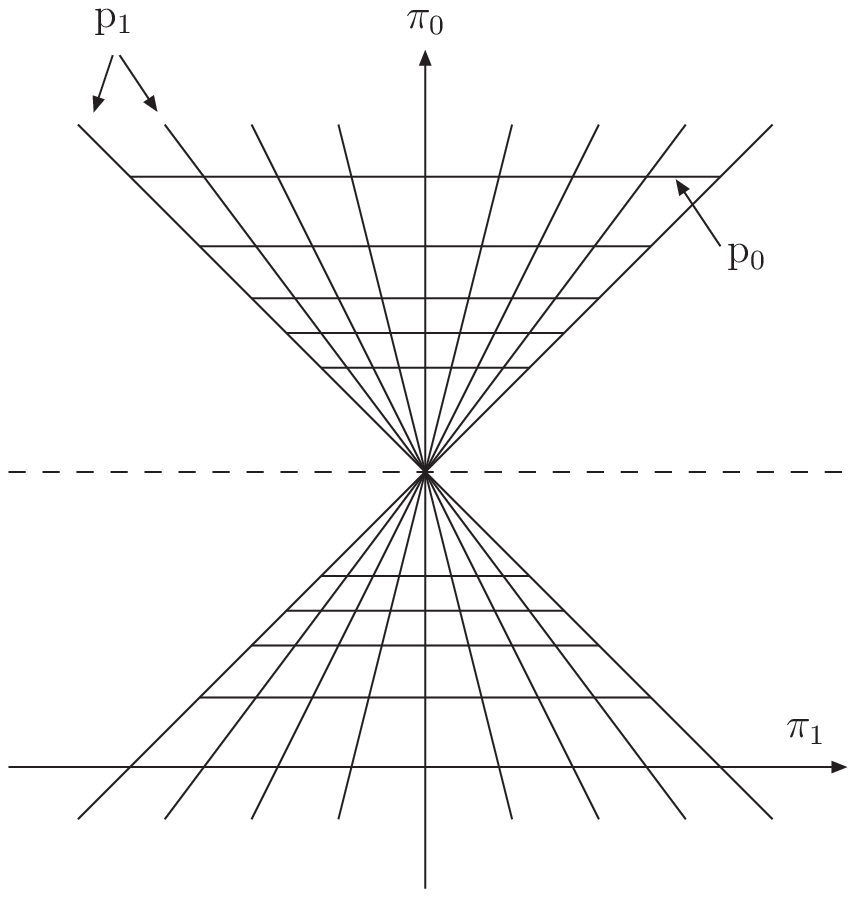}
\end{picture}
\end{center}

\caption{Magueijo - Smolin( MS)  spacetime  diagram in terms of
$\pi$ coordinates. ~~~~~MS coordinates $p_{0}$ and $p_{1}$ is drawn
as horizontal and radial lines.}
\end{figure}

\section{Amelino-Camelia DSR}
The other main example of DRS theories is the Amelino-Camelia DSR.
Amelino-Camelia was the first physicist that wanted to solve the
mentioned puzzle: \emph{How could $l_{p}$ play a role in the
structure of spacetime without violating Special Relativity.} He
modified the basic postulates of the Einstein's Relativity as

 1. The laws of physics involve a fundamental velocity scale "$c$" and a fundamental length
 "$l_{p}$".

 2. Each inertial obsrever can establish the value of $l_{p}$ (same value for all
 inertial observer) by determining the dispersion relation for photons, which takes
  the form $E^2-c^2p^2+f(E,p;l_{p})=0$, where $f$ is the same for all inertial observers
  and in particular all inertial observers agree on the leading $l_{p}$ dependence of
 $f$: $f(E,p;l_{p})\simeq l_{p}cp^{2}E$.

 If we find that the dispersion relation takes the form of

 \begin{eqnarray}
 2E^2[ \cosh{(\frac{E}{E_{p}})}-\cosh(\frac{m}{E_{p}})]=\vec{p}^{~2} e^{E/E_{p}}
 \end{eqnarray}

 by some reasoning or by the experimental analysis.

 The next step will be finding the deformed boost
 transformations which leave the above dispersion relation as an
 invariant.

 In ordinary special relativity the boosts can be described by

 \begin{eqnarray}\label{eq10}
  B_{a}=ip_{a}\frac{\partial}{\partial E}+ iE\frac{\partial}{\partial p_{a}}
 \end{eqnarray}

 By assuming that the modified generators also obey the same
Lorentz algebra and they preserve the dispersion relation
(\ref{eq10}) as an invariant, he finds the following modified
generators of boosts
 \begin{eqnarray}\label{eq11}
  B_{a}=ip_{a}\frac{\partial}{\partial E}+i\left(\frac{1}{2E_{p}}\vec{p}^{~2}+
  E_{p}\frac{1-e^{-2E/E_{p}}}{2}\right)\frac{\partial}{\partial p_{a}}-
  i\frac{p_{a}}{E_{p}}\left(p_{b}\frac{\partial}{\partial
  p_{b}}\right).
 \end{eqnarray}

 Please, note that the rotation generators remain intact. One can
 easily obtain the finite boost transformations that relate the
 observations of two observers by integrating the familiar
 differential equations
    $$ \frac{d E}{d\xi}=i[B_{a},E],~~~~\frac{dp_{b}}{d\xi}=i[B_{a},p_{b}].$$

But, here as in the Magueijo-Smolin transformations case we can see
that by defining the new variables $\epsilon$ an $\pi$ through the
relations

 \begin{eqnarray}\label{eq12}
  \frac{\epsilon}{\mu}=\frac{e^{E/E_{p}}-\cosh{(m/E_{p})}}{\sinh(m/E_{p})},~~~~~~~
  \frac{\pi}{\mu}=\frac{p~e^{E/E_{p}}}{E_{p}\sinh(m/E_{p})}
 \end{eqnarray}
 all relations come back to the ordinary special
 relativity forms. For example, the modified boost will
 take the form of usual Lorentz boost:
$$B_{a}=i\pi_{a}\frac{\partial}{\partial \epsilon}+
i\epsilon\frac{\partial}{\partial \pi_{a}}.$$

Note that the transformations (\ref{eq12}) are non-singular and we
can define the variables $\epsilon$ and $\pi$.

So, it seems that the Amelino-Camelia DSR like the MS ones is only a
re-description of the special relativity in the non-conventual
coordinates.
\section{Conlcusion}
From the above discussion it seems that Doubly Special theories are
only re-descriptions of the special relativity in the non-Cartesian
and non-conventional coordinates.

~~~~~~~~~~~~~~~~~~~~~~~~~~~~~~~~~~~~~~~~~~~~~~~~~~~~~~~~~~~~~~~~~~~~~~~~~~~~~~~~~~~~~
~~~~~~~~~~~~~~~~~~~~~~~~~~~~~~~~~~~~~~~~~~~~~~~~~~~~~~~~~~~~~~~~~~~~~~~~~~~~~~~~~~~~~
~~~~~~~~~~~~~~~~~~~~~~~~~~~~~~~~~~~~~~~~~~~~~~~~~~~~~~~~~~~~~~~~~~~~~~~~~~~~~~~~~~~~~
~~~~~~~~~~~~~~~~~~~~~~~~~~~~~~~~~~~~~~~~~~~~~~~~~~~~~~~~~~~~~~~~~~~~~~~~~~~~~~~~~~~~~

\end{document}